\documentclass[superscriptaddress,amsmath,amssymb,aps,twocolumn,floatfix]{revtex4-1}
\usepackage[english]{babel}
\usepackage[colorinlistoftodos]{todonotes}
\usepackage{multirow}
\usepackage{comment}
\usepackage{relsize}
\usepackage{float}
\usepackage{textcomp}
\usepackage[colorlinks=true,allcolors=.,urlcolor=blue]{hyperref}
\usepackage[justification=justified]{subcaption}
\usepackage[justification=justified, figurename=FIG.]{caption}
\usepackage[compact]{titlesec}
\titlespacing{\section}{0pt}{3ex}{2ex}
\titlespacing{\subsection}{0pt}{2ex}{2ex}
\titlespacing{\subsubsection}{0pt}{1.5ex}{1.5ex}

\DeclareCaptionLabelFormat{UC}{\MakeUppercase{#1} #2}
\usepackage{xcolor}
\usepackage{graphicx}
\usepackage{epstopdf}
\usepackage{color}
\usepackage{dcolumn}
\usepackage{bm}
\usepackage{hyperref}
\usepackage[mathlines]{lineno}
\usepackage{threeparttable}
\usepackage{booktabs} 
\usepackage{siunitx}
\usepackage{placeins}
\usepackage{xspace}

\newcommand\Tstrut{\rule{0pt}{2.6ex}}         

\begin{document}

\title{First measurement of discrimination between helium and electron recoils in liquid xenon for low-mass dark matter searches}

\author{S.J.~Haselschwardt} \email[Corresponding author: ]{scotthaselschwardt@lbl.gov}
\affiliation{Lawrence Berkeley National Laboratory, 1 Cyclotron Road, Berkeley, CA 94720, USA}
\author{R.~Gibbons}
\affiliation{Lawrence Berkeley National Laboratory, 1 Cyclotron Road, Berkeley, CA 94720, USA}
\affiliation{University of California, Berkeley, Department of Physics, Berkeley, CA 94720-7300, USA}
\author{H.~Chen}
\affiliation{Lawrence Berkeley National Laboratory, 1 Cyclotron Road, Berkeley, CA 94720, USA}
\author{S.~Kravitz} \thanks{Now at University of Texas at Austin, Austin, TX 78712, USA}
\affiliation{Lawrence Berkeley National Laboratory, 1 Cyclotron Road, Berkeley, CA 94720, USA}
\author{A.~Manalaysay}
\affiliation{Lawrence Berkeley National Laboratory, 1 Cyclotron Road, Berkeley, CA 94720, USA}
\author{Q.~Xia}
\affiliation{Lawrence Berkeley National Laboratory, 1 Cyclotron Road, Berkeley, CA 94720, USA}
\author{W.H.~Lippincott}
\affiliation{University of California, Santa Barbara, Department of Physics, Santa Barbara, CA 93106-9530, USA}
\author{P.~Sorensen}
\affiliation{Lawrence Berkeley National Laboratory, 1 Cyclotron Road, Berkeley, CA 94720, USA}

\date{\today}
\begin{abstract}
\noindent
We report the first measurement of discrimination between low-energy helium recoils and electron recoils in liquid xenon. This result is relevant to proposed low-mass dark matter searches which seek to dissolve light target nuclei in the active volume of liquid-xenon time projection chambers. Low-energy helium recoils were produced by degrading $\alpha$ particles from $^{210}$Po with a gold foil situated on the cathode of a liquid xenon time-projection chamber. The resulting population of helium recoil events is well separated from electron recoils and is also offset from the expected position of xenon nuclear recoil events.

\end{abstract}

\maketitle

\emph{Introduction --} 
Searches for interactions of galactic dark matter particles in terrestrial experiments span a diverse suite of detection techniques. Experiments seeking elastic scattering of dark matter particles on xenon nuclei have set the most stringent constraints on interactions of dark matter with mass  $\gtrsim$5~GeV/$c^2$, led by the dual-phase, time projection chamber (TPC) technology~\cite{LUX:2016ggv,XENON:2018voc,PandaX-4T:2021bab,LZ_1st_res,XnT_1st_res}. While this mass range is well motivated by candidates such as the weakly interacting massive particle, there is significant interest to also search for lower-mass dark matter.

Elastic scattering of $\lesssim$5~GeV/$c^2$ particles on xenon results in recoils below the few-keV threshold of modern xenon TPC detectors. A potential method for extending the sensitivity of these experiments to $\mathcal{O}(1)$~GeV/$c^2$ dark matter particles is to dissolve light nuclei, such as hydrogen or helium, into the xenon volume, as in HydroX~\cite{Lippincott:2017yst}. 
This provides a target with better kinematic match to the light dark matter projectile, resulting in recoils above the detection threshold through excitation and ionization of the xenon medium.

A crucial feature of xenon-based experiments is their ability to discriminate between the xenon nuclear recoils (NRs) expected from dark matter interactions and the background of electron recoils (ERs) from radioactivity and solar neutrinos. 
In dual-phase TPCs this discrimination is enabled by the simultaneous measurement of recoil ionization electrons and scintillation photons. 
NR events produce, on average, less ionization signal than ER events for a given amount of scintillation, a result influenced by both the initial ratio of excited/ionized atoms in the track and the fraction of ionized electrons which promptly recombine.
While the background suppression provided through ER/NR discrimination in liquid-xenon (LXe) TPCs is now well calibrated~\cite{LUX:2020car}, it is not known if such an effect is present for recoils of dissolved light nuclei.

In this paper we describe a measurement of $\gtrsim$~2~keV helium recoils in LXe using a custom, low-energy $\alpha$-particle source. 
This is the first measure of LXe signal response to NRs from $Z<54$ nuclei. A comprehensive general treatment of the energy transfer of a recoiling projectile (with atomic number $Z_1$) to motion of recoiling atoms (with atomic number $Z_2$) and to atomic electrons was given by Lindhard et al.~\cite{Lindhard:1963abc}. The theoretical treatment is complex for the case of $Z_1=Z_2$, and for $Z_1 \neq Z_2$ only an outline of the treatment was given. Our data are the first to inform this theoretical treatment for the case in which LXe comprises the target medium. We observe significant discrimination between helium recoil events and background ER events. 
Our results suggest for the first time that rejection of ER backgrounds is possible in light dark matter searches using LXe doped with light nuclei and motivate further characterization of this effect.

\begin{figure}[t]
\center
\includegraphics[width=\columnwidth]{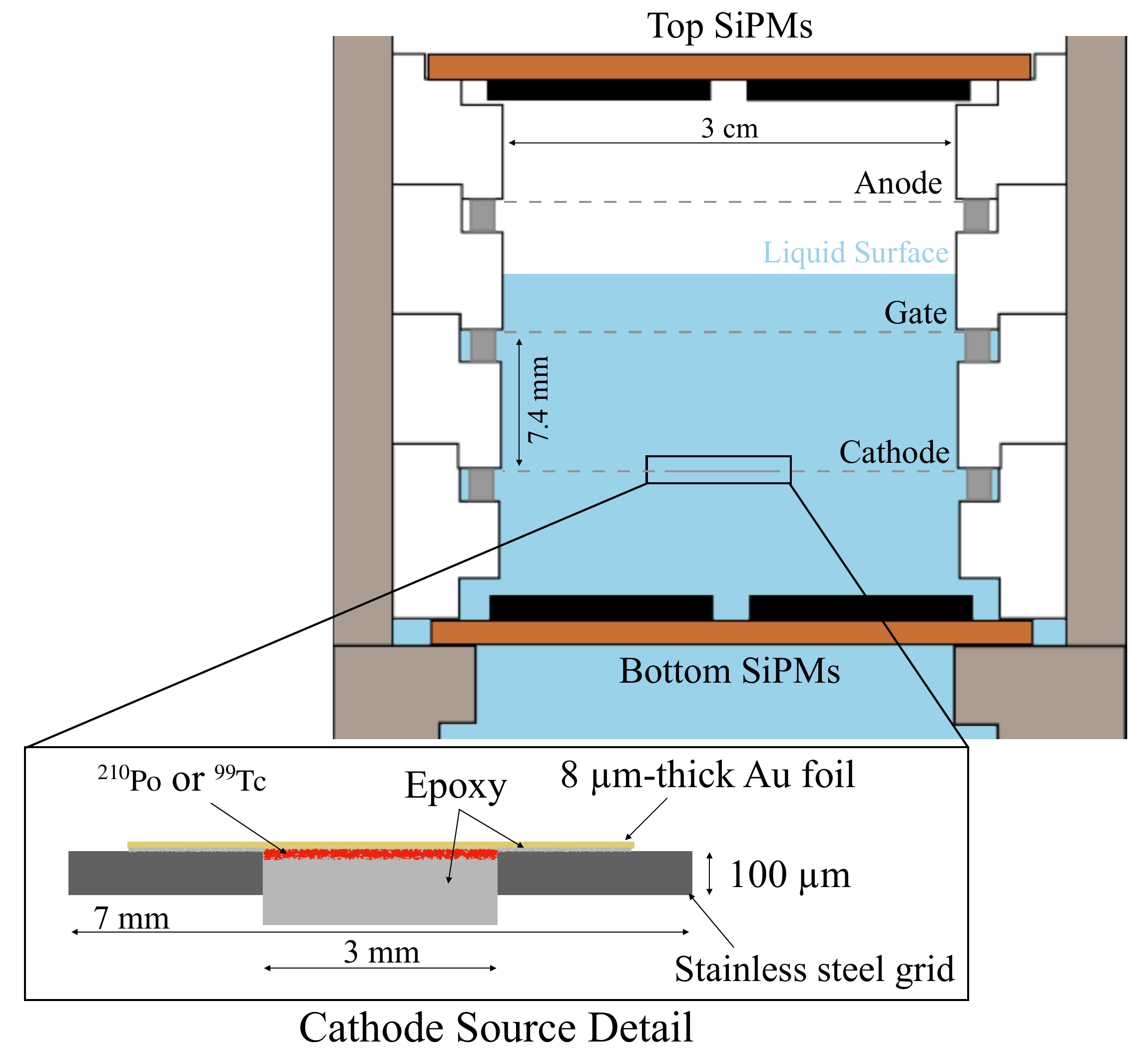}
\caption{Schematic cross sectional view of the dual-phase TPC. The inset shows details of the cathode source described in the text.}
\label{fig:expt}
\end{figure}

\emph{Experimental setup and data collection --}
The dual-phase TPC and associated cryogenic system used in these measurements is an upgrade of the detector described in~\cite{Kravitz:2022mby}, shown in Fig.~\ref{fig:expt}. A cylinder with 3~cm inner diameter PTFE walls and three electrostatic grids defines the principle region of the detector. The active xenon volume with height 0.74~cm is bounded from below and above by the cathode and gate grids, respectively. The anode grid is situated 0.74~cm above the gate, and the xenon liquid-vapor interface lies in the region between them. Each grid is a hexagonal stainless steel mesh with 2.7~mm pitch and wire thickness 100~$\mu$m. The cathode grid additionally features the source of low-energy $\alpha$ particles described in detail below. For the data reported here the cathode, gate, and anode voltages were held at -5.2~kV, -5.0~kV, and 0.5~kV, respectively.
 
The xenon vapor pressure was maintained at 1.5~bar with a stability of 5\%. The system was baked and pumped to remove impurities from TPC components prior to cooling. The xenon gas was purified through a hot metal getter before condensation, and the vapor above the liquid was continuously circulated and purified during each run. No changes in detector performance were observed during the data collection periods.

Particle interactions in the active LXe produce scintillation photons and ionization electrons. The promptly detected scintillation photon signal is denoted ``S1". Ionization electrons are drifted upward in the electric field set by the cathode and gate voltages to the liquid-vapor boundary where they are extracted into the vapor by the much higher field set by the gate-anode voltage difference. Electrons in the vapor phase rapidly accelerate and create electroluminescence photons, the signal from which is denoted ``S2". 
The time separation between the S1 and S2 signals, or ``drift time", determines the depth of the interaction in the detector. The relative size of the S2 signal determines the nature of the primary interaction as S2s from ERs are characteristically larger than those from xenon NRs for a given S1 size.

Photons are detected using arrays of silicon photomultipliers (SiPMs) at the top and bottom. 
Each array is a $2\times2$ assembly of Hamamatsu S13371 devices, each of which contain four independent photosensors, making 32 total readout channels. 
Each photosensor has dimension 5.95~mm$\times$5.85~mm and pixel pitch of 50~$\mu$m. 
The distribution of S2 photons in the top array allows for reconstruction of each event's lateral $(x,y)$ position in the detector, calibrated using the known radius of the TPC wall. The SiPM signals are each amplified $\times2$ before being digitized at a sampling frequency of 500~MS/s and saved for offline pulse identification and processing. Signal sizes are reported in photoelectrons (pe) after correcting for the mean area of single-photon-induced pulses measured in each channel. This corrects for SiPM afterpulsing and internal crosstalk as described in~\cite{Gibbons:2023iux}.

Event readout was triggered on the coincidence of any two channels exceeding a 10~mV ($\approx$8~pe) threshold in a 16~ns window, a condition fully efficient for S2s from the low-energy $\alpha$ events of interest. The corresponding trigger rate from each source was roughly 7~Hz. S2s from single electrons extracted from the liquid phase have average, corrected pulse areas of $\approx$20~pe, varying by $<$3\% over the course of each run. Large S2s result in SiPM saturation, and the deviation from linearity  is expected to be roughly 7\% for detected pulse areas near $31,500$~pe.

\begin{figure}[t]
\center
\includegraphics[width=\columnwidth]{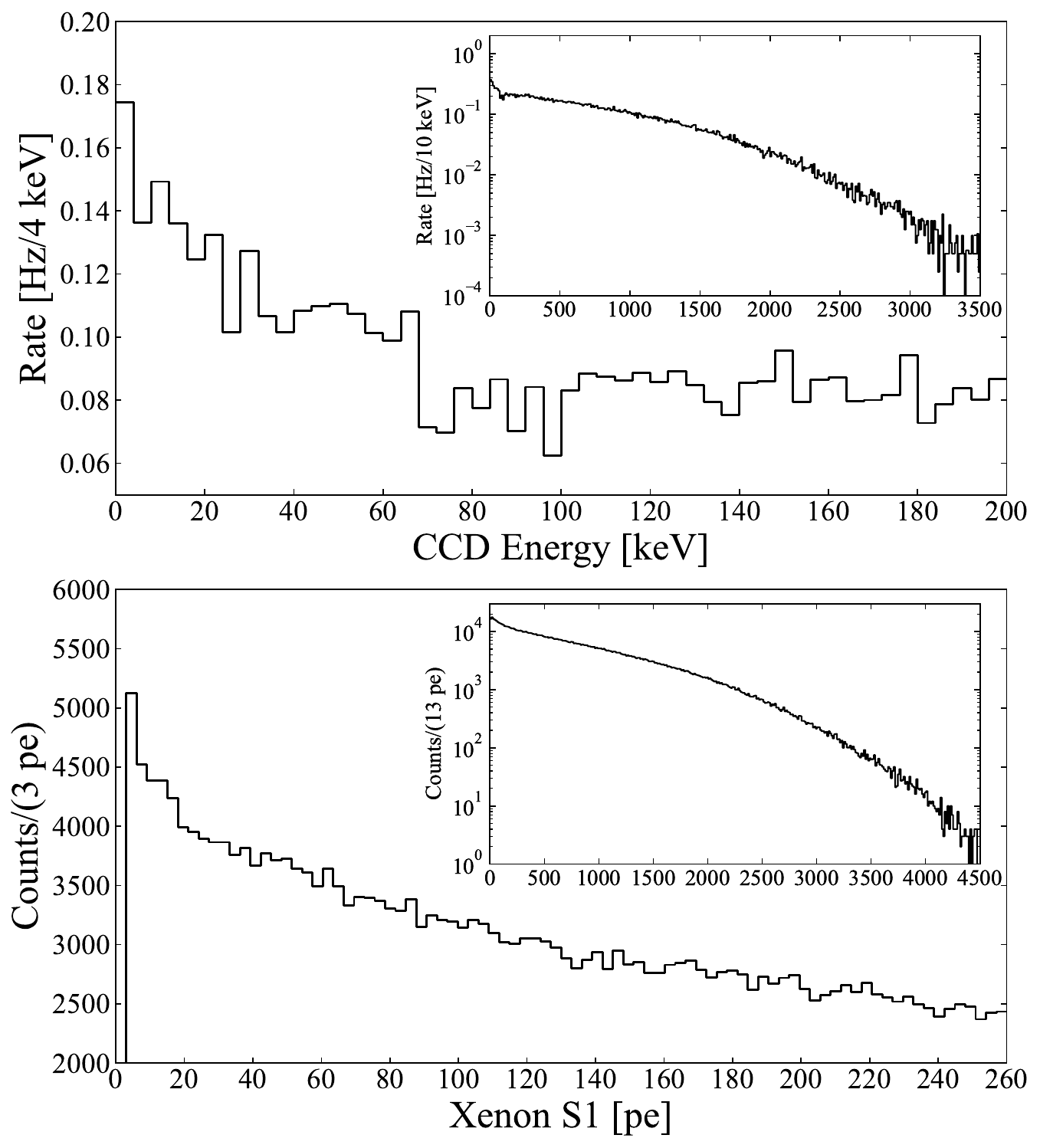}
\caption{\emph{Top:} The energy spectrum of particles emerging from the degraded $\alpha$ source measured in a back-illuminated CCD after subtracting the spectrum of cosmic-ray background. \emph{Bottom:} The S1 spectrum (roughly proportional to energy) of source events in the LXe TPC. Insets in both panels show the spectra over the full energy range.
}
\label{fig:energy}
\end{figure}

\begin{figure*}[t]
\center
\includegraphics[width=\textwidth]{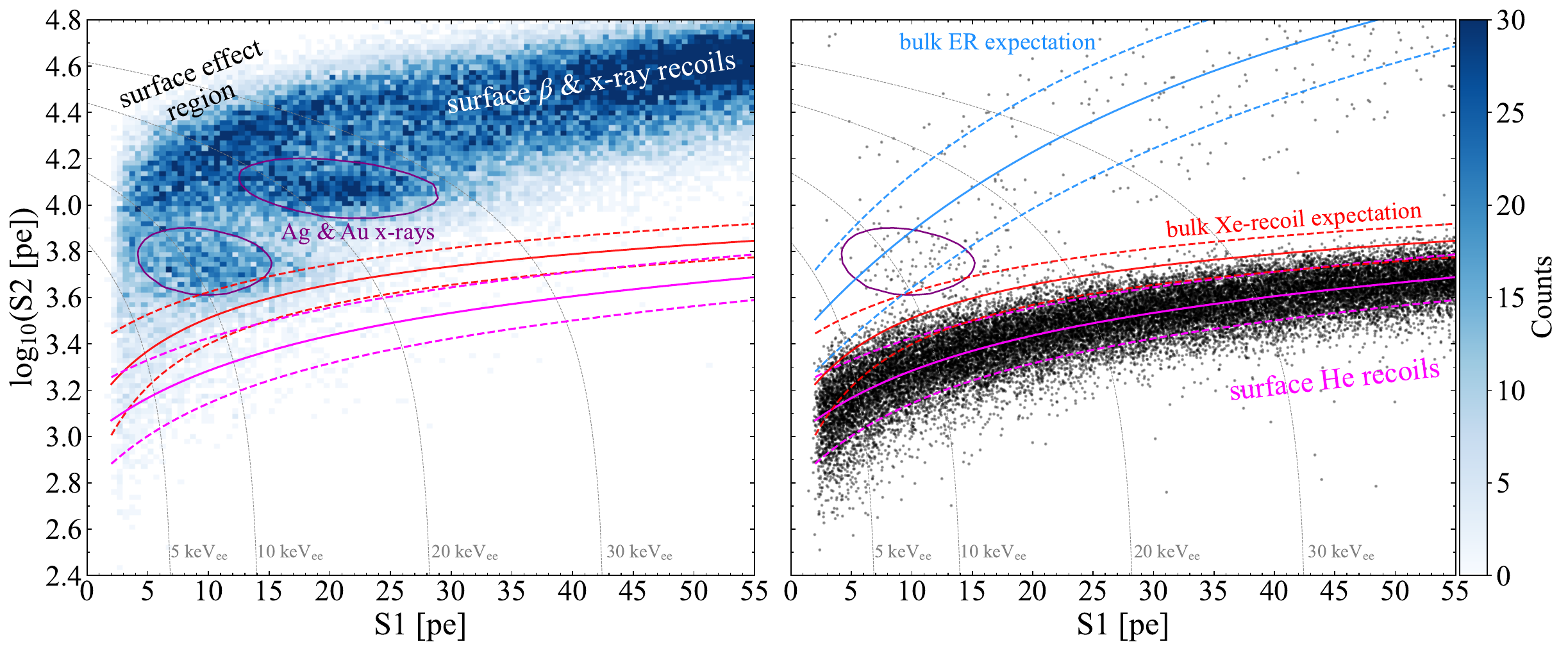} 
\caption{ \emph{Left:} The blue histogram shows the $^{99}$Tc $\beta$ calibration data. Purple ovals display contours containing 90\% of the gold and silver x-ray populations from the tuned detector response model.
\emph{Right:} Black points show events from the low-energy $\alpha$ source. For the sake of clarity, the plot shows 40\% of the total He-recoil data set. In both panels the solid (dashed) magenta lines define the median (10\% and 90\% boundaries) of the helium recoil band calculated from the data. Blue and red lines similarly characterize the bands expected from flat energy spectra of bulk ER and xenon NR events based on the response model, respectively.
Dashed grey curves denote contours of constant electron-equivalent energy (denoted keV$_{\rm ee}$). ERs from ambient $\gamma$ rays and gold x-rays are also visible in the $\alpha$ source data. The nonlinearity in detector response is $>7$\% for S2s larger than $10^{4.5}$~pe. Systematic effects that may influence the position of the He-recoil events are described in the text.}
\label{fig:discrim}
\end{figure*}

Low-energy helium recoils originate from a source at the center of the cathode, shown schematically in Fig.~\ref{fig:expt}. The center of the cathode features an annular ring in the mesh grid plane with inner and outer diameters of 3~mm and 7~mm, respectively. 
An 8~$\mu$m thick, 6~mm diameter gold foil disk is glued to the top of the annulus using cryogen-compatible, conductive epoxy. $^{210}$Po is deposited on the underside of the gold foil with spot size limited by the 3~mm inner diameter of the annulus ring to prevent interactions near the disk's edge where non-uniformity of the electric field is expected. The $^{210}$Po is then sealed in place with a 0.25~mm-thick layer of epoxy to prevent downward-going $\alpha$'s from interacting in the LXe below the cathode (where no accompanying S2 signal is collected).

Decays of $^{210}$Po produce 5.3~MeV $\alpha$ particles, roughly 3\% (as assessed in simulation) of which penetrate the gold foil after losing a significant fraction of their energy. 
The foil thickness of 8~$\mu$m (with 25\% uncertainty quoted by the manufacturer) produces a continuous distribution of $\alpha$ energies from an endpoint near 3.2~MeV down to zero~keV. 
The top panel of Fig.~\ref{fig:energy} shows the energy spectrum of $\alpha$'s from the source as characterized in a separate, back-illuminated charged-coupled device (CCD) test bed, similar to that in~\cite{Fernandez-Moroni:2020abn}. The CCD used here features inactive layers with total thickness $\approx$180~nm which blocks incident $\alpha$'s with energies $<20$~keV and further degrades the spectrum. The bottom panel of Fig.~\ref{fig:energy} shows the S1 spectrum of events from the source in the LXe TPC.

An identically prepared source containing only $^{99}$Tc was installed in a separate run to calibrate the detector response to low-energy ERs near the gold surface. $^{99}$Tc decays primarily to the ground state of $^{99}$Ru, emitting $\beta$'s with endpoint energy 297.5~keV. Simulations using Geant4~\cite{geant4} predict the spectrum of $\beta$'s that penetrate the gold foil has a reduced endpoint near 250~keV. Beta interactions with the gold foil and silver in the source epoxy result in the emission of x-rays near 10~keV (gold L-shell transitions) and 22~keV (silver K-shell transitions).

\emph{Results--}
Events containing only a single S1 pulse and its corresponding S2 pulse are included in our analysis. Coincidence of at least two SiPM channels is required of the S1 pulse to avoid the background of single-pe dark rate in the SiPMs. 
Events from each source are readily distinguished from ambient backgrounds using their drift times and reconstructed radial positions. Source events are required to have radius within 5~mm of the detector center and fall within a 0.3~$\mu$s-wide window centered on the TPC's maximum drift time ($\approx5$~$\mu$s).

Figure~\ref{fig:discrim} shows the detector response to the $\alpha$ and $\beta$ sources for events selected as described above. The S2 sizes in $\alpha$-source data are scaled by 0.9 to align the S2 peaks from gold x-rays present in both data sets.
A striking separation is observed between the ER and helium recoil event populations, in excess of the separation typically observed for recoiling xenon nuclei. 

The two x-ray peaks in the $\beta$ source data are used to tune a model for the detector response near the cathode based on the NEST~\cite{nest} simulation. 
The free parameters of the model are the electric field strength and the S1 photon detection efficiency. The best fit electric field strength is $780\pm140$~V/cm, in agreement with an electrostatic simulation of the TPC which uses a 50~$\mu$m finite-element spacing. 
The fit S1 detection efficiency is $0.020\pm0.002$~pe/photon, approximately half the value in the TPC center as expected from geometric shadowing of scintillation light by the source disk. The model uses an S2 detection efficiency of $19\pm4$~pe/electron, where the uncertainty stems from that in the xenon liquid level height. Purple contours in Fig.~\ref{fig:discrim} display quantiles of the x-ray populations from the best-fit response model. Blue and red lines show the model's predicted bands for bulk ERs and xenon NRs.

\emph{Discussion-- }A complication inherent in this measurement is the fact that many low-energy $\beta$ and $\alpha$ particles will have a range smaller than that of the gold and silver x-rays and are therefore sampling a region significantly closer to the source surface. The proximity of the surface to the events in question leads to several systematic uncertainties discussed below. Pursuant to this, Table~\ref{tab:range} gives the range in LXe of particles emitted from the two sources.

\begin{table}[t]
\caption{Range~\cite{astar,xcom} of particles ejected from the $\alpha$ and $\beta$ sources in LXe at various energies of interest. Projected (as opposed to continuous slowing down approximation) ranges are given for $\alpha$'s and $\beta$'s, while the $e^{-1}$ interaction length is given for x-rays.} \label{tab:range}
\begin{ruledtabular}
\begin{tabular}{cll}

Particle & Energy (keV) & Range ($\mu$m) \\
\hline
\Tstrut \multirow{2}{*}{x-ray} & 10 & 20 \\
 & 22 & 163 \\ [1ex]
\multirow{3}{*}{$\beta$} & 10 & 0.20 \\
& 62 & 4.5 \\
 & 250 & 45 \\[1ex]
 \multirow{3}{*}{$\alpha$} & 10 & 0.13 \\
 & 765 & 4.5 \\
 & 3000 & 19 \\

\end{tabular}
\end{ruledtabular}
\end{table}

Based on the preceding discussion, we expect three primary manifestations of surface effects, listed here in order of importance: (1) loss of ionized electrons to the metal surface, (2) surface electric field values higher than the 780~V/cm characterizing the x-ray populations, and (3) emission of secondary electrons from the surface in conjunction with the primary particle.

(1): $\alpha$ particles with energy $\lesssim765$~keV have projected ranges $\lesssim4.5$~$\mu$m, the calculated thermalization length of ionized electrons in LXe~\cite{Mozumder1995}. 
Therefore, ionized electrons from such particle tracks may be lost to the gold surface through attractive surface features such as the work function of the gold and/or the creation of image charges. 
We expect the charge loss probability to be roughly 50\% if the charge thermalization profile (from the quasi-linear alpha track) is spherically symmetric.


(2): The lowest-energy $\beta$ events in the 2--20~pe S1 region of Fig.~\ref{fig:discrim}, labeled ``surface effect region", have higher than expected S2 yields relative to the detector model tuned to the x-ray populations. This suggests that the low-energy $\beta$'s and $\alpha$'s experience much stronger electric fields in close proximity ($\lesssim1~\mu$m) to the gold surface (caused by, e.g., surface roughness). This may be caused by, for example, surface roughness, and we note here that the $\alpha$ and $\beta$ source foils may not have identical roughness profiles. 
Comparisons with NEST suggest that electric field strengths of $\sim5$~kV/cm could account for the high charge yield of these $\beta$ events. If one assumes the field dependence measured for $\sim5$~MeV $\alpha$'s~\cite{Jorg:2021hzu} holds for $\alpha$ energies $\lesssim50$~keV, we would expect the observed helium recoil ionization signal to be roughly 40\% larger than its value at the nominal $E=780$~V/cm. However, because these particle tracks are already in the Lindhard stopping regime~\cite{PDG} (on the low-energy side of the Bragg peak), we expect it more likely that the low-energy charge yield has little dependence on electric field strength, as is the case for recoiling xenon nuclei~\cite{Aprile:2006kx,Aprile:2018jvg}.

(3): The emission of secondary electrons from metals due to impinging and through-going heavy ions is well-known (see, e.g.,~\cite{se_review}, for a review). The electron yield generically depends on the electronic stopping power of the primary particle near the metal surface. For helium ions incident on gold, the vast majority of emitted electrons have energy below the 13.7~eV mean excitation energy of LXe (e.g.,~\cite{HASSELKAMP1986,Lohmann:2020oun}) such that they most likely thermalize and contribute to the total ionization signal of the primary interaction. The secondary electron yield accompanying emission of 3~MeV and 1~MeV $\alpha$'s from a gold foil was previously measured to be roughly 10 electrons~\cite{Anno1963,Yarger1966}, and more recent studies have found lower yields using lower-energy helium ions~\cite{Eder1997}. 
Ignoring the dependence on electronic stopping and assuming a secondary electron yield of 10 electrons, our measured ionization signal from 5~keV helium recoils would be enhanced by about 15\%. By similar reasoning the affect on the position of the ER band in $\beta$-source data is negligible.

Of these three effects, we expect only the first of charge loss to the foil surface to be substantial. This would act to increase the observed discrimination from ER events in our data. Interestingly, if half of the measured ionization signal from $\alpha$ tracks were lost to the surface, the actual band from recoiling helium nuclei in bulk LXe would nearly align with the known band from recoiling xenon nuclei. In this context, we note that an earlier version of the $\alpha$ source included $^{210}$Po on the top surface of the gold foil, in direct contact with the active xenon. This produced a source of 0--100~keV $^{206}$Pb recoils from decays in which the $\alpha$ was emitted into the foil. The charge yield band of those $^{206}$Pb recoils is nearly identical to that of the helium recoils, suggesting that a wide range of low-energy recoiling atomic species ($^4$He, $^{131}$Xe, $^{206}$Pb) produce a similar partitioning of deposited energy into scintillation and ionization. Such an effect is likely influenced by the resolution of the instrument, with higher-resolution leading to modest separation (see, e.g.,~\cite{Strauss:2014zia}).



We have made a first measurement of low-energy helium recoils in a LXe TPC and observed a clear separation from the ER band, a promising result for dark matter searches with LXe TPCs doped with a light target. 
We have shown for the first time that the band from helium-recoil events will likely coincide (within experimental resolution) to that from xenon NRs. 
Our work motivates more detailed studies of low-mass NRs in LXe, particularly the associated photon and ionization yields.
An ideal (but challenging to realize) experiment would measure the ionization and scintillation from recoiling helium nuclei dissolved in the bulk LXe.
If those recoils are caused by an external neutron source, then our work suggests such a measurement will face the irreducible systematic uncertainty that the recoiling light species and xenon populations will be nearly indistinguishable, though the endpoint of their recoil spectra will be vastly different. 

\emph{Acknowledgments --} We thank Henrique Araujo and Eric Dahl for helpful discussions. This work was supported by the Office of Science, Office of High Energy Physics, of the U.S. Department of Energy under Contract Nos. DE-AC02-05CH11231 and DE-SC0021115.

\FloatBarrier

\bibliographystyle{apsrev4-2}
\bibliography{main}

\end{document}